\documentclass[12pt]{article} \def\theequation{\arabic{section}.\arabic{equation}}
\newcounter{rown}

\usepackage[centertags]{amsmath}
\usepackage{color}
\usepackage{amsfonts}
\usepackage{amssymb}
\usepackage{amsthm}
\usepackage{newlfont}
\begin{document}
\renewcommand{\theequation}{\arabic{section}.\arabic{equation}}
\title{{\bf Extended $D=3$ Bargmann supergravity from a Lie algebra expansion}}
\vskip 4cm
\author{ J.A. de Azc\'{a}rraga, \\
Departamento de F\'{\i}sica Te\'orica and IFIC (CSIC-UVEG), \\
 46100-Burjassot (Valencia), Spain\\
 D. G\'utiez,\\
 Department of Physics, Universidad de Oviedo, \\
33007-Oviedo, Spain\\
 J. M. Izquierdo,\\
Departamento de F\'{\i}sica Te\'orica, Universidad de Valladolid, \\
47011-Valladolid, Spain}
\date{Jul. 12, 2019}
\maketitle
\vskip 1cm
\begin{abstract}
In this paper we show how the method of Lie algebra expansions may be used to obtain, in a simple way, both the
extended Bargmann Lie superalgebra and the Chern-Simons action associated to it in three dimensions, starting from
$D=3$, $\mathcal{N}=2$  superPoincar\'e and its corresponding Chern-Simons supergravity.
\end{abstract}
\newpage

\section{Introduction}

In recent years, the supersymmetric version of Newtonian Gravity, {\it i.e.} Newtonian  supergravity, has received some attention in the context of a non-relativistic version of the AdS/CFT correspondence (see, for instance, \cite{BG:09,CHOR:14}). However, certain problems need to be solved, and progress is still being made.

The natural way to address the problem requires using a Galilean superalgebra as a starting point, plus a gauging procedure that, in the bosonic case, should recover Newtonian gravity. In \cite{ABPR:11} this was done in $D=3,4$ and in the absence of fermions by starting from the centrally extended Galilei algebra or Bargmann algebra, and imposing certain conditions on the curvatures associated to its gauging. These conditions allowed the authors to obtain Newtonian gravity in the Newton-Cartan (NC) formalism \cite{C:23}, which formulates Newtonian gravity in a way that resembles general relativity. Subsequently, the supersymmetric case was studied \cite{ABRS:13,BRZ:15} in $D=3$ for a superalgebra that contains two fermionic generators and such that the bosonic part is the Bargmann algebra. In this way, the $D=3$ NC supergravity was obtained.

The solution to the problem mentioned above, however,
has some limitations: spatial geometry is fixed to be flat, and there is no satisfactory action principle associated to it.
 An action was obtained in \cite{BR:16} that overcomes these difficulties, which was called the extended Bargmann supergravity action. In contrast with the NC supergravity case, the bosonic part of the supersymmetry algebra is a further extension of the Bargmann algebra, and the action itself is a Chern-Simons (CS) action, as it is also the case of $D=3$ Poincar\'e supergravity \cite{AT:86}. In fact, it was shown in \cite{BR:16} that both the Galilei action and algebra may be obtained from the Poincar\'e ones as a limit that, although it looks like a contraction, it is not so since it does not preserve the dimension of the algebra.

In this paper, we point out that the Galilean superalgebra and the CS action mentioned above may be found alternatively by using the method of
Lie algebra expansions, which has its origin in the work of \cite{HS:03} and was formulated and studied in general in \cite{AIPV:03,AIPV:07} (see also \cite{AI:12,AILW:13} for other applications and \cite{IRS:06} for a generalization involving semigroups\footnote{For D=3 constructions based on the Nappi-Witten
and the AdS-Lorentz algebras see \cite{PS:19} and \cite{CR:19}, and
\cite{GRSS:16} for D = 5 starting from a CS gravity.}). In addition to the other three ways
of obtaining new algebras from given ones, namely contractions and deformations (both preserving the dimension of the original Lie algebra) and extensions (which require two algebras), the method of expansions provides a way to obtain in general larger Lie
algebras from a given one (see \cite{AIPV:07}). Presumably, all algebras obtained by the expansion method may be obtained by a combination of extensions and contractions, but from the computational point of view expansions are more interesting, as some calculations simplify considerably.

More precisely, we show here that, starting from $D=3$, $\mathcal{N}=2$ Poincar\'e supergravity and a CS action associated to it, the method of expansions applied to the algebra leads to the extended Bargmann superalgebra of \cite{BR:16}. Further, when the superPoincar\'e action is expanded, the result is also the action of \cite{BR:16}. As it will become apparent, the calculations involved are very simple.

The plan of the paper is the following. In Sec. 2 we will briefly review the method of Lie algebra expansions \cite{AIPV:03,AIPV:07} in the particular case of interest to us. In section 3 it will be applied to the $D=3$ Poincar\'e gravity to obtain a bosonic Galilean CS action. Section 4 is devoted to the expansions of the $D=3$, $\mathcal{N}=2$ superPoincar\'e based CS model  that leads to the algebra and action obtained in \cite{BR:16}. In the conclusions, we will comment on the possible future applications of the method in the context of Galilean gravity and supergravity.

\section{Lie Algebra expansions}

In a nutshell, Lie algebra expansions consists of three steps. Given a Lie (super)algebra $\mathcal{G}$,
\begin{enumerate}
  \item Write a formal series expansion in $\lambda$ of the Maurer-Cartan (MC)
  one-forms associated with the Lie algebra,
  \item Insert the expansions into the MC equations of $\mathcal{G}$ and identify equal powers of $\lambda$ to obtain a consistent
  infinite set of MC equations, and
  \item Cut the infinite expansions in a consistent way so that a finite Lie algebra, the expanded algebra, is obtained in terms
  of its MC equations.
\end{enumerate}

We give the details of the construction in the case of interest in this paper, namely when $\mathcal{G}$ has a
symmetric coset structure.
Let  $\mathcal{G}=V_0 \oplus V_1$, where $V_1$ is a symmetric coset,
\begin{equation}
\label{simcos}
   [V_0,V_0]\subset V_0 \ , \quad [V_0,V_1]\subset V_1\ ,\quad [V_1,V_1]\subset V_0\ .
\end{equation}
Let $\omega^{i_p}$, be the MC forms valued on the spaces  $V_p$, $p=0,1$
and $i_p=1,\dots , \mathrm{dim} V_p$. Let us write the MC equations of $\mathcal{G}$ in the form
\begin{equation}\
\label{MCsup} d \omega^{k_s} = - \frac{1}{2} c^{k_s}_{i_p j_q}
\omega^{i_p} \wedge \omega^{j_q} \quad (p,q,s= 0,1) \ .
\end{equation}
Condition \eqref{simcos} implies that the structure constants of the algebra satisfy
\begin{equation}\label{simcosb}
   c^{k_1}_{i_0 j_0}= 0 \ , \quad c^{k_0}_{i_0 j_1} = 0 \ , \quad c^{k_1}_{i_1 j_1}= 0 \ .
\end{equation}
Then, it is consistent to expand the MC forms of $V_0$ in
terms of even powers of $\lambda$ and those of $V_1$ in terms of
odd powers of $\lambda$ as
\begin{eqnarray}
\label{expthinf}
   \omega^{i_0} &=& \sum^{\infty}_{\alpha_0=0,\ \alpha_0 \, even}
\lambda^{\alpha_0} \omega^{i_0,\alpha_0} \nonumber \\
 \omega^{i_1} &=& \sum^{\infty}_{\alpha_1=1,\ \alpha_1 \, odd}
 \lambda^{\alpha_1} \omega^{i_1,\alpha_1}
\ .
\end{eqnarray}
When these expansions are inserted in the MC equations \eqref{MCsup}, and the equal powers of $\lambda$ in both sides are identified, we obtain a consistent infinite number of MC one-forms and equations. To obtain finite Lie algebras, the expansions must be cut in a consistent way, so that they correspond to the MC equations of a Lie (super)algebra. It can be shown that this is achieved when
\begin{eqnarray}
\label{expth}
   \omega^{i_0} &=& \sum^{N_0}_{\alpha_0=0,\ \alpha_0 \, even}
\lambda^{\alpha_0} \omega^{i_0,\alpha_0} \nonumber \\
 \omega^{i_1} &=& \sum^{N_1}_{\alpha_1=1,\ \alpha_1 \, odd}
 \lambda^{\alpha_1} \omega^{i_1,\alpha_1}
\ ,
\end{eqnarray}
provided that the $N_0$ and $N_1$ integers
satisfy one of the two conditions below
\begin{eqnarray}
\label{conditions}
  & & N_0= N_1+1 \, \nonumber\\
 & & N_0= N_1-1  \  ,
\end{eqnarray}
(see \cite{AIPV:03} for the proof). This leads to a series of finite-dimensional superalgebras, which
are denoted by $\mathcal{G}(N_0,N_1)$, with structure
constants given by
\begin{equation}
\label{expsc}
   C^{k_s,\alpha_s}_{i_p,\beta_p\, j_q,\gamma_q} =
\left\{\begin{array}{cc}
  0 & \mathrm{if}\; \beta_p+\gamma_q \neq \alpha_s \\
 c^{k_s}_{i_p j_q} & \mathrm{if}\; \beta_p+\gamma_q = \alpha_s\\
\end{array}\right.  \quad .
\end{equation}

From the MC equations we may obtain the gauge curvatures of the same Lie algebra by noticing that
the latter may be viewed as an equation that expresses the failure of the MC equations. So if $0=d \theta
+\theta \wedge \theta$ are the MC equations, then the curvatures are given by $F= dA + A \wedge A$,
and by taking $F=0$ we recover the MC equations\footnote{To avoid complicating the notation, in
this paper we will use the same symbols to denote the MC one-forms and the corresponding gauge one-form
fields.}.
Then, making the replacement
$\omega^{i_s,\alpha_s} \rightarrow A^{i_s,\alpha_s}$, the MC forms and MC equations are replaced by gauge
one-forms and by the equations defining of the curvatures,
\begin{equation}
\label{Curvatures}
F^{k_s,\alpha_s} = dA^{k_s,\alpha_s}+ \frac{1}{2}
C^{k_s,\alpha_s}_{i_p,\beta_p\, j_q,\gamma_q} A^{i_p,\beta_p}
\wedge
A^{j_q,\gamma_q} \ ,
\end{equation}
so that the MC equations may be recovered by setting $F^{k_s,\alpha_s}=0$.The Bianchi identities and  gauge variations (of infinitesimal
parameters $\varphi^{i_s,\alpha_s}$) are given by
\begin{eqnarray}
\label{expFDA}
d F^{k_s,\alpha_s} &=&  C^{k_s,\alpha_s}_{i_p,\beta_p\,
j_q,\gamma_q} F^{i_p,\beta_p} \wedge
A^{j_q,\gamma_q} \ , \nonumber\\
\delta A^{k_s,\alpha_s} &=& d \varphi^{k_s,\alpha_s} -
C^{k_s,\alpha_s}_{i_p,\beta_p\, j_q,\gamma_q}
\varphi^{i_p,\beta_p}  A^{j_q,\gamma_q} \ .
\end{eqnarray}
It is crucial for our construction to note that, alternatively, these equations may be obtained by substituting, in the
equations that would correspond to the original $\mathcal{G}$, that is ({\it cf.} \eqref{MCsup})
\begin{eqnarray}
\label{gaugeorig}
  F^{k_s} &=& dA^{k_s}+ \frac{1}{2}
c^{k_s}_{i_p\, j_q} A^{i_p}
\wedge
A^{j_q}  \ , \nonumber \\
\delta A^{k_s} &=& d \varphi^{k_s} -
C^{k_s}_{i_p\, j_q}
\varphi^{i_p} \wedge A^{j_q} \ ,
\end{eqnarray}
the expansions of $A^{k_s}$, $F^{k_s}$ and $\varphi^{k_s}$ with
exactly the same structure as the of $\omega^{k_s}$ in \eqref{expth} and then identifying equal powers
of $\lambda$ (see \cite{AIPV:03}).

\subsection{Expanded CS actions}

We can use the expansions of the gauge one-forms and curvature two-forms to obtain, in some cases, new actions from a given one.
As an example, we consider now the important case of the CS actions.

Let $\mathcal{G}$ be a Lie
superalgebra,  and let $k_{I_1,\dots I_l}$ be the coordinates of a
symmetric invariant
$l$-tensor of $\mathcal{G}$. Then, the
$2l$-form
\begin{equation}
\label{2lform}
   H= k_{I_1,\dots I_l} F^{I_1} \wedge \dots \wedge F^{I_l}
\end{equation}
is closed and invariant under gauge transformations. Since the
gauge FDAs (given by the definition of the curvatures plus the Bianchi identities) are contractible, this defines a $(2l-1)$-form $B$
(the CS form, see {\it e.g.} \cite{AI:95}), such that $dB=H$, and if $B$ is integrated over a
$(2l-1)$-dimensional manifold $\mathcal{M}^{2l-1}$, a CS model is
obtained through the action
\begin{equation}
\label{CSaction}
  I[A] = \int_{\mathcal{M}^{2l-1}} B(A) \ ,
\end{equation}
where ${\mathcal{M}^{2l-1}}$ is the $(2l-1)$-dimensional spacetime.

New CS actions for the expanded algebras may be obtained by inserting the expansions of $A^I$ and $F^I$
in the CS action for $\mathcal{G}$,
\begin{equation}
\label{CSexpansion}
 I[A,\lambda] = \int_{\mathcal{M}^{2l-1}} B(A,\lambda) =
 \int_{\mathcal{M}^{2l-1}} \sum_{N=0}^\infty \lambda^N B_N(A) =
\sum_{N=0}^\infty \lambda^N I_N[A] \ .
\end{equation}
The same expansion, when applied to
\eqref{2lform}, leads to
\begin{equation}
\label{Hexp}
  H(F,\lambda) = \sum_{N=0}^\infty \lambda^N H_N \ ,
\quad
   H_N = d B_N(A) \ .
\end{equation}
This means that the actions given by
\begin{equation}
\label{NCS}
  I_N = \int_{\mathcal{M}^{2l-1}} B_N(A)
\end{equation}
define  CS models that have been obtained
by expanding of the original $\mathcal{G}$-based one. The  corresponding
Lie algebra is the smallest one that contains {\it all} the fields appearing in $I_N$. Not keeping all the fields may
result in a lack of gauge invariance of the actions, which is otherwise guaranteed if the expansion is kept infinite. Then,
in practice,  the power of $\lambda$ in the expansion of the action selects the corresponding finite expanded algebra.
In general, the expanded actions and algebras `remember' the structure of the original ones (see Eq. \eqref{expsc}) a fact
that simplifies the calculations.

One computational advantage of expansions is the fact that the
equation of motion for $A^{k_s,\alpha_s}$ in $I_N$, which may be represented by $E(A^{k_s,\alpha_s})=0$, satisfies
\begin{equation}
      E(A^{k_s,\alpha_s})= E(A^{k_s})|_{N-\alpha_s} ,
\end{equation}
where $E(A^{k_s})|_{N-\alpha_s}$ is the coefficient of $\lambda^{N-\alpha_s}$ in
the expansion of $E(A^{k_s})$.

\section{Galilei expansion of arbitrary $D$ poincar\'e and $D=3$ gravity.}

Before going to the supersymmetric case, we consider in this section the bosonic expanded algebras and action
to illustrate the method. Although the subject of this paper is $D=3$, we will keep $D$ arbitrary for the expansion
of the gauge fields and curvatures, and fix $D=3$ when constructing the action (gravity in $D>3$ is not CS; see,
however, the Outlook).

\subsection{Poincar\'e algebra and space-time splitting}

Our starting algebra $\mathcal{G}$ will be the Poincar\'e algebra in arbitrary dimensions, which in a certain basis can be described by the MC equations
\begin{eqnarray}\label{MCPoin}
d \tilde{e}^A &=& - \tilde{\omega}^A{}_B \wedge \tilde{e}^B \nonumber\\
d \tilde{\omega}^{AB} &=& -\tilde{\omega}^A{}_C \wedge \tilde{\omega}^{CB} \ ,
\end{eqnarray}
where $A,B,C=0,\dots , D-1$. We will use a `mostly plus' $(1,D-1)$ signature for the Minkowski metric $\eta_{AB}$.
In order to perform an expansion leading to an extension of the Galilei algebra, we split the Poincar\'e algebra
generators as follows:
\begin{eqnarray}
\label{bosonicspl}
\tilde{e}^A & \rightarrow & ( \tilde{e}^a, \, \tilde{e}^0=\tilde{\phi}) \ , \nonumber\\
\tilde{\omega}^{AB} & \rightarrow & ( \tilde{\omega}^{ab},\,  \tilde{\omega}^a{}_0= \tilde{\omega}^a)\ ,
\end{eqnarray}
where $a=1,\dots , D-1$. In terms of these one-forms, the MC equations read
\begin{eqnarray}
\label{MCPoinSp}
  d \tilde{e}^a &=& - \tilde{\omega}^a{}_b \wedge \tilde{e}^b - \tilde{\omega}^a \wedge \tilde{\phi} \nonumber\\
  d \tilde{\phi} &=& - \tilde{\omega}_a \wedge \tilde{e}^a \nonumber\\
  d \tilde{\omega}^{ab} &=& -\tilde{\omega}^a{}_c \wedge \tilde{\omega}^{cb} - \tilde{\omega}^a \wedge \tilde{\omega}^b \nonumber\\
  d \tilde{\omega}^a &=& -\tilde{\omega}^a{}_b \wedge  \tilde{\omega}^b \ .\end{eqnarray}

As mentioned earlier, the gauge curvatures can be viewed as the two-forms that express the failure of the MC equations;
then, the MC one-forms become the gauge one-form fields (again, denoted by the same symbols).
The gauge curvatures of the Poincar\'e algebra are the two-forms $\widetilde{T}^A$, $\widetilde{R}^{AB}$ given by
\begin{eqnarray}
\label{GaugePoin}
  \widetilde{T}^A &=& d \tilde{e}^A + \tilde{\omega}^A{}_B \wedge \tilde{e}^B \\
  \widetilde{R}^{AB} &=& d \tilde{\omega}^{AB} + \tilde{\omega}^A{}_C \wedge \tilde{\omega}^{CB}\ .
\end{eqnarray}
By using the space-time splitting for the curvatures,
\begin{eqnarray}
\label{bosonicsplG}
\widetilde{T}^A & \rightarrow & ( \widetilde{T}^a, \, \widetilde{T}^0=\widetilde{\Omega}) \ , \nonumber\\
\widetilde{R}^{AB} & \rightarrow & ( \widetilde{R}^{ab},\,  \widetilde{R}^a{}_0= \widetilde{R}^a)\ ,
\end{eqnarray}
the gauge curvatures $\widetilde{T}^a$, $\widetilde{\Omega}$, $\widetilde{R}^{ab}$ and $\widetilde{R}^a$ are given in terms of the gauge fields by
\begin{eqnarray}
\label{GaugePoinSp}
 \widetilde{T}^a &=& d \tilde{e}^a  + \tilde{\omega}^a{}_b \wedge \tilde{e}^b + \tilde{\omega}^a \wedge \tilde{\phi} \nonumber\\
  \widetilde{\Omega}  &=& d \tilde{\phi} - \tilde{\omega}_a \wedge \tilde{e}^a \nonumber\\
  \widetilde{R}^{ab} &=& d \tilde{\omega}^{ab}  +\tilde{\omega}^a{}_c \wedge \tilde{\omega}^{cb} + \tilde{\omega}^a \wedge \tilde{\omega}^b \nonumber\\
  \widetilde{R}^a &=& d \tilde{\omega}^a  +\tilde{\omega}^a{}_b \wedge \tilde{\omega}^b \ .
\end{eqnarray}
It is seen that the MC equations \eqref{MCPoin} and \eqref{MCPoinSp} are recovered when the curvatures are set to zero in \eqref{GaugePoin} and \eqref{GaugePoinSp}.

\subsection{Expansion of the algebra and the $D=3$ action}

If we choose $V_0^*$ as the vector space generated by $\tilde{\omega}^{ab}, \tilde{\phi}$, and $V_1^*$ as the
one generated by $\tilde{e}^a, \tilde{\omega}^a$, we have precisely the structure \eqref{simcos}. Thus, we may perform
the following consistent expansion in terms of a parameter $\lambda$:
\begin{equation}
\label{BosExp}
\begin{array}{ll}
  \tilde{e}^a= \lambda e^a + \sum_{k=1}^\infty \lambda^{2k+1} \tilde{e}^a_{(2k+1)}\, ,  &
   \widetilde{T}^a =\lambda T^a + \sum_{k=1}^\infty \lambda^{2k+1} \widetilde{T}^a_{(2k+1)} \\
  \tilde{\phi} = \phi +\lambda^2 \varphi + \sum_{k=2}^\infty \lambda^{2k} \tilde{\phi}_{(2k)}\, , & \widetilde{\Omega} = \Omega + \lambda^2 \Lambda + \sum_{k=2}^\infty \lambda^{2k} \widetilde{\Omega}_{(2k)} \\
  \tilde{\omega}^{ab} = {\omega}^{ab}+ \lambda^2 \ell^{ab} + \sum_{k=2}^\infty \lambda^{2k} \tilde{\omega}^{ab}_{(2k)} \, , & \widetilde{R}^{ab} = R^{ab} + \lambda^2 L^{ab} + \sum_{k=2}^\infty \lambda^{2k} \widetilde{R}^{ab}_{(2k)}\\
  \tilde{\omega}^a = \lambda \omega^a + \sum_{k=1}^\infty \lambda^{2k+1} \tilde{\omega}^a_{(2k+1)}\, , &
  \widetilde{R}^a= \lambda R^a
  + \sum_{k=1}^\infty \lambda^{2k+1} \widetilde{R}^a_{(2k+1)}
\end{array}\ .
\end{equation}
The expansion is infinite, but it may be cut in a consistent manner (see eq. \eqref{conditions}).
As argued before, we will consider the finite algebra that contains all the fields that appear in a suitable term of the
expanded action. More explicitly, let us start from the four-form $\widetilde{H}$ given by
\begin{eqnarray}
\label{tildeHB}
  \widetilde{H} = \epsilon_{ABC} \widetilde{R}^{AB} \wedge \widetilde{T}^{C}  ,
\end{eqnarray}
with $A,B,C = 0,1,2$. This form is closed, $d\widetilde{H} = 0$, so there exists a three-form $\widetilde{B}$ such that $d\widetilde{B}=\widetilde{H}$.
The integral over three-dimensional spacetime gives an action that describes general relativity in three
dimensions and in the absence of matter. We will however use $\widetilde{H}$, because it is much simpler to
derive the field equations from it, and also exhibits the CS character of the action.
Let us now rewrite $\widetilde{H}$ using the space-time splitting \eqref{bosonicspl} and \eqref{bosonicsplG}:
\begin{equation}
\label{tildeHBspl}
 \widetilde{H} = 2 \epsilon_{ab} \widetilde{R}^{a} \wedge \widetilde{T}^{b}  + \epsilon_{ab} \widetilde{R}^{ab}\wedge \widetilde{\Omega}\ ,
\end{equation}
where $\epsilon_{ab}$ is the Levy-Civita symbol in $2$ dimensions, $\epsilon_{0 ab} = \epsilon_{ab}$.

Let us now replace the fields in \eqref{tildeHBspl} by  their expansions \eqref{BosExp}. This leads to
an expansion of $\widetilde{H}$,
\begin{equation}\label{expH}
    \widetilde{H}  =  \sum_{k=0}^{\infty} \lambda^k \widetilde{H}|_k \ ,\end{equation}
where the terms $\widetilde{H}|_k$ depend on the fields of the expansion and, since they are closed, define actions on
these fields. The gauge algebra corresponding to a particular term $\widetilde{H}|_k$ will be the consistent truncation
of the infinite expansion that has the gauge fields corresponding to the curvatures that it contains.

The lowest order term in $\lambda$ of the expansion of the first term of \eqref{tildeHBspl} is $\lambda^{2} \epsilon_{ab} R^{a} \wedge T^{b} $. We need to keep this term if the resulting action has to be related with gravity, because we need $T^a=0$ and the contribution for the $\omega^a$ equation of this term will be of this sort. This means that our model corresponds to the term $\widetilde{H}|_{2}$ in \eqref{expH}. We now have to find out which curvatures appear in the $4$-form $\widetilde{H}$ of \eqref{tildeHBspl} .
By selecting the $\lambda^2$ in the expansion of $\widetilde{H}$, we obtain the four form  $H$  given by
\begin{equation}
\label{HD3B}
H= \widetilde{H}|_2= 2\epsilon_{a_1 a_2} R^{a_1} \wedge T^{a_2}  + \epsilon_{a_1 a_2} L^{a_1a_2} \wedge \Omega + \epsilon_{a_1 a_2} R^{a_1a_2} \wedge \Lambda  \ .
\end{equation}
which means that the gauge curvatures for  this model are
\begin{eqnarray}
\label{gaugeB2}
  T^a &=& de^a +\omega^a{}_b\wedge e^b \nonumber + \omega^a \wedge \phi \\  \Omega &=& d\phi \nonumber \\
  \Lambda &=& d\varphi + \omega_a \wedge e^a \nonumber \\
  R^{ab} &=& d\omega^{ab} +\omega^a{}_c \wedge \omega^{cb} \nonumber \\
  L^{ab} &=& d\ell^{ab} +\omega^a{}_c \wedge \ell^{cb} + \ell^a{}_c \wedge \omega^{cb} +\omega^a\wedge \omega^b\nonumber \\
  R^a &=& d\omega^a+ \omega^a{}_b \wedge \omega^b \ ,
\end{eqnarray}
expressions that are valid for any $D$. For $D=3$, $R^{ab}$ reduces to $d\omega^{ab}$ and
the second and third terms in the expression of $L^{ab}$ cancel each other.
The corresponding Lie algebra  (remember that the MC equations may be recovered by setting the curvatures to zero)
is precisely the extension of the Bargmann algebra studied by Bergshoeff et al. in \cite{BR:16}. If we set the
curvatures equal to zero,  the MC forms $e^a$ are dual to the generators of space translations, $\phi$ is dual to
the generator of the time translations, $\omega_a$ correspond to the Galilean boosts, and $\omega_{ab}$  to the
rotations, while $\varphi$ and $\ell^{ab}$ are dual to commuting extension generators that determine,
respectively, the Bargmann and the extended Bargmann algebra.
Note that \eqref{HD3B} is closed and only depends on the curvatures. Hence $H$ is invariant under the gauge
transformations of the algebra corresponding to \eqref{gaugeB2}, and therefore defines a CS action, which coincides with
the bosonic sector of the one obtained in \cite{BR:16}.

\subsubsection{On the physical dimensions of $\lambda$}

We now comment on the issue of the physical dimensions of the expansion parameter $\lambda$.  Although the expansion in terms of
powers of the parameter $\lambda$ is formal,  the gauge fields ultimately involved have physical dimensions. This is
achieved in general by assigning a suitable dimension to the parameter $\lambda$. In \cite{AIPV:03}, $D=3$ Poincar\'e supergravity
was obtained by expanding a CS action based on a simple superalgebra. The generators of a simple algebra are dimensionless,
and those of the superPoincar\'e algebra have to be dimensionful if they are to be associated with Poincar\'e supergravity,
so $\lambda$ has to have dimensions. In our case, the starting Poincar\'e fields do have dimensions, but these are
different from the dimensions of the fields in the expansion \eqref{gaugeB2}.

Let us start with the fields in the Poincar\'e action. We may choose $[\tilde{e}^A]=T$, while  $\omega^{AB}$ has
to be dimensionless. Since the  metric is given in terms of the dreibein  by\begin{equation}
\label{eg}
       g_{\mu\nu} = e^A_\mu e_{A\nu}\ ,
\end{equation}
where $e^A= e^A_\mu dx^\mu$ ($e^A_\mu$ are the coordinates of $e^A$ in the basis $dx^\mu$). If we take $[x^0]=T$, $[x^a]=L$,
then $g_{00}$ is dimensionless and $[g_{ij}]=T^2L^{-2}$. This is compatible with the flat spacetime expression
\begin{equation}
\label{mink}
       \left(
         \begin{array}{ccc}
           -1 & 0 & 0  \\
           0 & \frac{1}{c^2} & 0  \\
           0 & 0 & \frac{1}{c^2}  \\
         \end{array}
       \right) \ .
\end{equation}
Now, consider the algebra obtained by setting the curvatures equal to zero in \eqref{gaugeB2}. Since $\omega^a$
correspond to the Galilean boosts, it makes sense to set $[\omega^a]=L T^{-1}$. Also, we would like to have $[e]=L^{-1}T$
as they are dual to the space translations. From \eqref{BosExp} we deduce that $[\lambda]= L^{-1}T$, that is, the inverse
of a velocity. This argument, of course, is valid in every dimension $D$. This means that it is consistent, although not
necessary in this context, to assume that the parameter $\lambda$ is equal to $c^{-1}$. For a construction that does
include a $c^{-1}$ expansion, see \cite{HHO:19}.

\section{From $N=2$ superPoincar\'e to the extended su\-per\-Galilei in $D=3$}

We now start from the superPoincar\'e algebra in $D=3$, which in terms of its MC forms is given by
\begin{eqnarray}
\label{sPD3}
d \widetilde{e}^A &=& -\widetilde{\omega}^A{}_B \wedge \widetilde{e}^B -i \overline{\widetilde{\psi}}
\gamma^A \wedge \widetilde{\psi} \nonumber\\
d \widetilde{\omega}^{AB} &=& -\widetilde{\omega}^A{}_C \wedge \widetilde{\omega}^{CB} \nonumber\\
d \widetilde{\psi} &=& - \frac{1}{4} \gamma^{AB} \widetilde{\omega}_{AB} \wedge \widetilde{\psi} \ ,
\end{eqnarray}
where $A,B,C=0,1,2$ and we are using the $(-++)$ metric (we use the convention that complex conjugation
reorders the product of Grassmann-odd symbols).

When the fermion one-forms $\widetilde{\psi}$ are complex, the algebra is that of $\mathcal{N}=2$ superPoincar\'e,
and it is $\mathcal{N}=1$ when they are Majorana spinors. We are interested in obtaining a superGalilei algebra by expanding
superPoincar\'e, with anticommutators of the type $\{ Q, Q\} \propto H$, where $Q$ is a $so(2)$ spinor supersymmetry
generator and $H$ generates the time translations. Looking at the most general expansion it turns out that this requires
starting from $\mathcal{N}=2$ superPoincar\'e. Additionally, this fact may be justified by noticing that we
will need to split the original $so(1,2)$ spinor into two $so(2)$ spinors as suggested by the results of \cite{ABRS:13},
but a real  $so(1,2)$ spinor has two real components, so the $so(2)$ spinors must have one real component each. But this
would correspond to Majorana-Weyl spinors, which do not exist in $D=2$ with signature $(++)$ (although they do exist when
the signature is $(-+)$). So we are forced to consider the case $\mathcal{N}=2$. In what follows our spinors will be
complex, with no reality condition assumed.

Let us now perform the space-time splitting including the fermions. First,  we take $\gamma^A$ real for convenience;
 for instance,
\begin{equation}\label{D3realgamma}
\gamma^0=i\sigma^2\ ,\quad \gamma^1=\sigma^1 \ ,\quad \gamma^2=\sigma^3 \ .
\end{equation}
We then define the following one-forms:
\begin{eqnarray}
\label{fermionicspl}
\tilde{e}^A & \rightarrow & ( \tilde{e}^a, \, \tilde{e}^0=\tilde{\phi}) \ , \nonumber\\
\tilde{\omega}^{AB} & \rightarrow & ( \tilde{\omega}^{ab},\,  \tilde{\omega}^a{}_0= \tilde{\omega}^a)\nonumber\\
\tilde{\psi} &=& P_+ \tilde{\xi}_+ + P_- \tilde{\xi}_- \ \quad \left( \overline{\tilde{\psi}} =  \overline{\tilde{\xi}_+} P_+ +  \overline{\tilde{\xi}_-} P_-\right)\ ,
\end{eqnarray}
where $P_\pm= \frac{1}{2}(1\pm i\gamma_0)$, and $\tilde{\xi}_\pm$ are real, as can be seen from
\begin{equation}
\label{lambdapsi}
\tilde{\xi}_\pm = \textrm{Re}\, \tilde{\psi} \pm \gamma^0 \textrm{Im}\, \tilde{\psi} \ .
\end{equation}
In terms of these forms, the MC equations \eqref{sPD3} read
 \begin{eqnarray}
\label{MCspPoinSp}
  d \tilde{e}^a &=& - \tilde{\omega}^a{}_b \wedge \tilde{e}^b - \tilde{\omega}^a \wedge \tilde{\phi}-i \overline{\tilde{\xi}}_+ \gamma^a \wedge \tilde{\xi}_- \nonumber\\
  d \tilde{\phi} &=& - \tilde{\omega}_a \wedge \tilde{e}^a + \frac{i}{2} \tilde{\xi}^t_+   \wedge \tilde{\xi}_+ + \frac{i}{2} \tilde{\xi}^t_-   \wedge \tilde{\xi}_-\nonumber\\
  d \tilde{\omega}^{ab} &=& -\tilde{\omega}^a{}_c \wedge \tilde{\omega}^{cb} - \tilde{\omega}^a \wedge \tilde{\omega}^b \nonumber\\
  d \tilde{\omega}^a &=& -\tilde{\omega}^a{}_b \wedge  \tilde{\omega}^b\nonumber\\
  d \tilde{\xi}_\pm &=& -\frac{1}{4} \omega_{ab} \gamma^{ab}\wedge \tilde{\xi}_\pm -\frac{1}{2}
  \gamma^a \tilde{\omega}_a \gamma^0 \wedge \tilde{\xi}_\mp \ .
\end{eqnarray}
As before, the MC one-forms become the gauge one-forms (denoted by the same letters), and the
gauge curvatures of the Poincar\'e algebra are the two-forms $\widetilde{T}^A$, $\widetilde{R}^{AB}$, $\tilde{\rho}$ given by
\begin{eqnarray}
\label{GaugesPoin}
  \widetilde{T}^A &=& d \tilde{e}^A + \tilde{\omega}^A{}_B \wedge \tilde{e}^B +i \overline{\tilde{\psi}} \gamma^A \wedge \tilde{\psi} \nonumber \\
  \widetilde{R}^{AB} &=& d \tilde{\omega}^{AB} + \tilde{\omega}^A{}_C \wedge \tilde{\omega}^{CB}\nonumber
  \\
  \tilde{\rho} &=& d\tilde{\psi} +\frac{1}{4} \tilde{\omega}_{AB}\gamma^{AB}\wedge \tilde{\psi} \ .
\end{eqnarray}
By using the space-time splitting for the curvatures,
\begin{eqnarray}
\label{fermionicsplG}
\widetilde{T}^A & \rightarrow & ( \widetilde{T}^a, \, \widetilde{T}^0=\widetilde{\Omega}) \ , \nonumber\\
\widetilde{R}^{AB} & \rightarrow & ( \widetilde{R}^{ab},\,  \widetilde{R}^a{}_0= \widetilde{R}^a)\ ,\nonumber\\
\tilde{\rho}&=& P_+ \widetilde{\Xi}_+ + P_- \widetilde{\Xi}_- \ ,
\end{eqnarray}
the gauge curvatures $\widetilde{T}^a$, $\widetilde{\Omega}$, $\widetilde{R}^{ab}$, $\widetilde{R}^a$ and $\widetilde{\Xi}_\pm$
are given in terms of the gauge fields by
\begin{eqnarray}
\label{GaugePoinSp1}
 \widetilde{T}^a &=& d \tilde{e}^a  + \tilde{\omega}^a{}_b \wedge \tilde{e}^b + \tilde{\omega}^a \wedge \tilde{\phi} + i \overline{\tilde{\xi}}_+ \gamma^a \wedge \tilde{\xi}_- \nonumber\\
  \widetilde{\Omega}&=&  d \tilde{\phi}  + \tilde{\omega}_a \wedge \tilde{e}^a - \frac{i}{2} \tilde{\xi}^t_+   \wedge \tilde{\xi}_+ - \frac{i}{2} \tilde{\xi}^t_- \wedge \tilde{\xi}_-\nonumber\\
  \widetilde{R}^{ab} &=& d \tilde{\omega}^{ab}  +\tilde{\omega}^a{}_c \wedge \tilde{\omega}^{cb} + \tilde{\omega}^a \wedge \tilde{\omega}^b \nonumber\\
  \widetilde{R}^a &=& d \tilde{\omega}^a  +\tilde{\omega}^a{}_b \wedge \tilde{\omega}^b\nonumber\\
  \widetilde{\Xi}_\pm &=& d\tilde{\xi}_\pm  +\frac{1}{4} \omega_{ab} \gamma^{ab}\wedge \tilde{\xi}_\pm +\frac{1}{2}
  \gamma^a \tilde{\omega}_a \gamma^0 \wedge \tilde{\xi}_\mp  \ .
\end{eqnarray}

We now expand the gauge one-forms and curvature two-forms contained in the gauge algebra \eqref{GaugePoinSp1}. To do this,
we notice that if we make the choice $V_0^*=\{\tilde{\omega}^{ab},\tilde{\phi},\tilde{\xi}_+\}$ and
$V_1^*=\{\tilde{e}^a, \tilde{\omega}^a, \tilde{\xi}_-\}$, then the superalgebra has the structure \eqref{simcos}.
So we may write
\begin{equation}
\label{FermExp}
\begin{array}{ll}
  \tilde{e}^a= \lambda e^a +  \sum_{k=1}^\infty \lambda^{2k+1} \tilde{e}^a_{(2k+1)}\, ,  &
   \widetilde{T}^a =\lambda T^a + \sum_{k=1}^\infty \lambda^{2k+1} \widetilde{T}^a_{(2k+1)} \\
  \tilde{\phi} = \phi +\lambda^2 \varphi + \sum_{k=2}^\infty \lambda^{2k} \tilde{\phi}_{(2k)}\, , & \widetilde{\Omega} = \Omega + \lambda^2 \Lambda + \sum_{k=2}^\infty \lambda^{2k} \widetilde{\Omega}_{(2k)} \\
  \tilde{\omega}^{ab} = {\omega}^{ab}+ \lambda^2 \ell^{ab} + \sum_{k=2}^\infty \lambda^{2k} \tilde{\omega}^{ab}_{(2k)} \, , & \widetilde{R}^{ab} = R^{ab} + \lambda^2 L^{ab} + \sum_{k=2}^\infty \lambda^{2k} \widetilde{R}^{ab}_{(2k)}\\
  \tilde{\omega}^a = \lambda \omega^a + \sum_{k=1}^\infty \lambda^{2k+1} \tilde{\omega}^a_{(2k+1)}\, , &
  \widetilde{R}^a= R^a
  + \sum_{k=1}^\infty \lambda^{2k+1} \widetilde{R}^a_{(2k+1)}\\
  \tilde{\xi}_+=  \psi +\lambda^2 \xi + \sum_{k=2}^\infty \lambda^{2k} \tilde{\xi}_+{}_{(2k)}\, , &
  \widetilde{\Xi}_+ = \rho + \lambda^2 \Xi + \sum_{k=2}^\infty \lambda^{2k} \widetilde{\Xi}_+{}_{(2k)}\\
  \tilde{\xi}_-= \lambda \pi + \sum_{k=1}^\infty \lambda^{2k+1} \tilde{\xi}_-{}_{(2k+1)}\, , &
  \widetilde{\Xi}_- = \lambda \Pi +  \sum_{k=1}^\infty \lambda^{2k+1} \widetilde{\Xi}_-{}_{(2k)}
\end{array}\ .
\end{equation}
Since we need to select the $\lambda^2$ term in the expansion of the $\mathcal{N}=2$, $D=3$ supergravity action,
we shall consistently cut the expansion at the power $\lambda^2$. The resulting gauge algebra is given by
\begin{eqnarray}
\label{gaugeF2}
  T^a &=& de^a +\omega^a{}_b\wedge e^b+ \omega^a \wedge \phi +i \bar{\psi} \gamma^a\wedge \pi\nonumber \\
  \Omega &=& d\phi -\frac{i}{2} \psi^t \wedge \psi \nonumber \\
  \Lambda &=& d\varphi + \omega_a \wedge e^a -i\psi^t\wedge \xi - \frac{i}{2} \pi^t \wedge \pi\nonumber \\
  R^{ab} &=& d\omega^{ab} +\omega^a{}_c \wedge \omega^{cb} \nonumber \\
  L^{ab} &=& d\ell^{ab} +\omega^a{}_c \wedge \ell^{cb} + \ell^a{}_c \wedge \omega^{cb} +\omega^a\wedge \omega^b\nonumber \\
  R^a &=& d\omega^a+ \omega^a{}_b \wedge \omega^b \nonumber\\
  \rho &=& d\psi + \frac{1}{4} \omega_{ab}\gamma^{ab} \wedge \psi\nonumber\\
  \Xi &=& d\xi + \frac{1}{4} \omega_{ab}\gamma^{ab} \wedge \xi  + \frac{1}{4} \ell_{ab}\gamma^{ab} \wedge \psi +
  \frac{1}{2} \gamma^a \omega_a\gamma^0\wedge \pi\nonumber\\
  \Pi &=& d\pi + \frac{1}{4} \omega_{ab}\gamma^{ab} \wedge \pi + \frac{1}{2} \gamma^a \omega_a\gamma^0\wedge \psi\ .
\end{eqnarray}
Again, the MC equations of the algebra are recovered setting all curvatures equal to zero. The bosonic subalgebra
is precisely the extended bosonic Bargmann algebra of \eqref{gaugeB2}. Eqs, \eqref{gaugeF2} contain also three real $so(2)$ odd
spinor gauge one-forms $\psi$, $\xi$ and $\pi$ and three fermionic curvatures $\rho, \Xi$ and $\Pi$.

\subsection{Dual version of the algebra}

Let us find the (anti)commutators of the generators dual to the MC forms of the algebra obtained from \eqref{gaugeF2}, when the curvatures vanish. Since in our $D=3$ case $a=1,2$, in order to make contact with \cite{ABRS:13} we write $\omega_{ab}=\epsilon_{ab}\omega$, $\ell_{ab}=\epsilon_{ab}q$ so the MC equations read
\begin{eqnarray}
\label{MCF2}
    de^a &=& -\epsilon^a{}_b\omega \wedge e^b- \omega^a \wedge \phi -i \bar{\psi} \gamma^a\wedge \pi\nonumber \\
    d\phi &=&- \frac{i}{2} \psi^t \wedge \psi \nonumber \\
    d\varphi &=& -\omega_a \wedge e^a +i\psi^t\wedge \xi + \frac{i}{2} \pi^t \wedge \pi\nonumber \\
  d\omega &=& 0 \nonumber \\
   dq &=& -\frac{1}{2}\epsilon_{ab} \omega^a\wedge \omega^b\nonumber \\
    d\omega^a&=& -\epsilon^a{}_b \omega\wedge \omega^b \nonumber\\
    d\psi &=& - \frac{1}{2} \omega \gamma^{0} \wedge \psi\nonumber\\
    d\xi &=& - \frac{1}{2} \omega\gamma^{0} \wedge \xi  - \frac{1}{2} q\gamma^{0} \wedge \psi
  -\frac{1}{2} \gamma^a \omega_a\gamma^0\wedge \pi\nonumber\\
   d\pi &=&   \frac{1}{2} \omega\gamma^{0} \wedge \pi + \frac{1}{2} \gamma^a \omega_a\gamma^0\wedge \psi\ .
\end{eqnarray}
Now we call the generators dual to  $e^a$, $\phi$, $\varphi$, $\omega$, $q$, $\omega^a$, $\psi$, $\xi$ y $\pi$,
$P_a$, $H$, $M$, $J$, $S$, $G^a$, $Q^+$, $U$ and $Q^-$ respectively. A convenient way of finding the commutators
is using of the canonical one-form
\begin{equation}
\label{canonical1}
   \theta = e^a P_a +\phi H + \varphi M+ \omega J+ q S + \omega^a G_a+ \psi^\alpha Q^+_\alpha + \xi^\alpha U_\alpha + \pi^\alpha Q^+_\alpha\ .
\end{equation}
In terms of $\theta$ the MC equations  $d\theta= -\theta \wedge \theta$ lead immediately to the superalgebra commutators
simply by inserting  $\theta= \omega^I X_I$. In this way, the following non-zero (anti)-commutators are obtained:
\begin{eqnarray}
\label{explie}
& & \left[G_a,H \right] =P_a \ ,\quad \left[G_a, P_b\right]= \delta_{ab} M \ ,\quad
\left[G_a, G_b\right] =\epsilon_{ab} S\ ,\nonumber\\
& &\left\{Q_\alpha^+, Q_\beta^+ \right\}= i\delta_{\alpha\beta}H \ ,\quad
 \left\{Q^+_\alpha, Q^-_\beta\right\}= -i(\gamma^0\gamma^a)_{\alpha\beta} P_a \ ,\nonumber\\
 & &  \{ Q^-_\alpha, Q^-_\beta\}= -i \delta_{\alpha\beta}M \ ,\quad \left\{Q^+_\alpha, U_\beta \right\} =-i \delta_{\alpha\beta}M \ ,\nonumber\\
& & \left[S, Q^+_\alpha\right] = \frac{1}{2}(\gamma^0)^\beta{}_\alpha U_\beta\ ,\quad
\left[G_a, Q^+_\alpha\right] = \frac{1}{2}(\gamma^0\gamma^a)^\beta{}_\alpha Q^-_\beta \ ,\nonumber\\
& & \left[G_a, Q^-_\alpha\right] = \frac{1}{2}(\gamma^0\gamma^a)^\beta{}_\alpha U_\beta \ ,\nonumber\\
& &  \left[J, P_a\right]=-\epsilon_{ab} P^b \ ,\quad \left[ J, G_a\right]= -\epsilon_{ab} G^b \ ,\nonumber\\
& & \left[J, Q^{\pm}_\alpha\right]= \frac{1}{2}(\gamma^0)^\beta{}_\alpha Q_\beta^{\pm} \ , \quad
 \left[J, U_\alpha\right] = \frac{1}{2}(\gamma^0)^\beta{}_\alpha U_\beta\ .\end{eqnarray}

The last two lines exhibit the semidirect structure of the algebra, $J$ being the generator of the two-dimensional rotations.
 The first line is the extended Bargmann algebra (omitting rotations), where $S$ is the generator of the central extension; the second
 and third lines contain the anticommutators of the fermionic generators and the fourth and the fifth one give the
 commutators of bosonic and fermionic generators, excluding the rotations.
Apart from minor redefinitions, these commutators coincide with those of \cite{BR:16}.

\subsection{Expansion of the action}

The next step is to expand the action, or equivalently $\widetilde{H}$, of $D=3$, $\mathcal{N}=2$ supergravity and select the coefficient of the $\lambda^2$ term. To this end, we need to start with the action of D=3 Poincar\'e supergravity. It is given by
\begin{equation}
\label{Hdef}
   \widetilde{H}= \epsilon^{ABC} \widetilde{R}_{AB}\wedge \widetilde{T}_C-4i \overline{\tilde{\rho}}\wedge \tilde{\rho}\ ,
\end{equation}
where $\overline{\tilde{\rho}}$ is the Dirac adjoint of $\tilde{\rho}$.
In order to check that the four-form $\widetilde{H}$, which depends only on the curvatures is closed and thus defines a CS action,
we have used that, with the choice \eqref{D3realgamma} of gamma matrices,  $\gamma^0\gamma^1\gamma^2= i\sigma^2\sigma^1\sigma^3=I_3$.
Thus, if we define $\epsilon^{012}=1$, then $\gamma^{ABC}=\epsilon^{ABC}$. This in turn gives
\begin{equation}\label{gammaabc}
   \gamma^{AB}= \epsilon^{ABC} \gamma_C \ .
\end{equation}
When calculating the exterior differential of $\widetilde{H}$, we have used the expression of the differentials of the curvatures
$\widetilde{R}_{AB}$, $\widetilde{T}_C$ and $\tilde{\rho}$,
\begin{eqnarray}\label{cFPoin}
  d\widetilde{R}_{AB} &=& \widetilde{R}_{AC}\wedge \tilde{\omega}^C{}_A-
  \tilde{\omega}_{AC}\wedge \widetilde{R}^C{}_A\ ,\quad (D\widetilde{R}_{AB} =0) \nonumber\\
  d\widetilde{T}^A &=& \widetilde{R}^A{}_B \wedge \tilde{e}^B - \tilde{\omega} ^A{}_B \wedge \widetilde{T}^B + i \overline{\tilde{\rho}} \gamma^A\wedge \tilde{\psi}- i\overline{\tilde{\psi}} \gamma^A\wedge \tilde{\rho} \ ,\nonumber\\
  & & \quad\quad\quad  (D\widetilde{T}^A = \widetilde{R}^A{}_B \wedge e^B  + i \overline{\tilde{\rho}} \gamma^A\wedge \tilde{\psi}- i\overline{\tilde{\psi}} \gamma^A\wedge \tilde{\rho})\nonumber\\
  d \tilde{\rho} &=& \frac{1}{4} \widetilde{R}_{AB} \gamma^{AB}\wedge \tilde{\psi} - \frac{1}{4} \tilde{\omega}_{AB} \gamma^{AB}\wedge \tilde{\rho}\ ,\nonumber\\
  && \quad\quad\quad (D \tilde{\rho} = \frac{1}{4} \widetilde{R}_{AB} \gamma^{AB}\wedge \tilde{\psi}\ ,\  D \overline{\tilde{\rho}} = \frac{1}{4} \overline{\tilde{\psi}} \wedge \widetilde{R}_{AB} \gamma^{AB}) \ ,
\end{eqnarray}
where $D$ is the Lorentz covariant exterior differential. Using Eqs. \eqref{cFPoin} and the gamma matrix identity \eqref{gammaabc}, the differential of $\widetilde{H}$ in \eqref{Hdef} is seen to vanish.

We have to rewrite \eqref{Hdef} in terms of the spacetime splitting \eqref{fermionicsplG} to perform the expansion. The result is (we write $\epsilon_{0ab}=\epsilon_{ab}$)
\begin{equation}
\label{Hdefspl}
   \widetilde{H}=2 \epsilon_{ab} \widetilde{R}^a\wedge \widetilde{T}^b + \epsilon_{ab} \widetilde{R}^{ab}
   \wedge \widetilde{\Omega} -2i \overline{\tilde{\Xi}}_+\wedge \tilde{\Xi}_+
   -2i\overline{\tilde{\Xi}}_-\wedge \tilde{\Xi}_- \ .
\end{equation}
We now expand the gauge one-forms and curvature two-forms as in \eqref{FermExp} and insert the expansion into \eqref{Hdefspl}.
Then, we select the $\lambda^2$ term, which is given by
\begin{eqnarray}
\label{H2ferm}
   H=\widetilde{H}|_2 & =& 2\epsilon_{ab} R^a\wedge T^b + \epsilon_{ab} R^{ab} \wedge \Lambda +\epsilon_{ab} L^{ab} \wedge \Omega \nonumber\\
  & &  -4i \bar{\rho} \wedge \Xi -2i \widetilde{\Pi} \wedge \Pi \ .
\end{eqnarray}
The equations of the action $I=\int_{\mathcal{M}^{3}} B$, where $dB=H$ and $\mathcal{M}^{3}$ denotes the two-dimensional
space and time, are given by the vanishing of all the curvatures included in \eqref{H2ferm}, and, since $H$
is gauge invariant under the gauge transformations that correspond to the gauge algebra \eqref{gaugeF2}, then $I$
will be invariant too, up to topological effects. The action \eqref{H2ferm} obtained coincides with that of \cite{BR:16}

\section{Outlook}

In this paper we have shown how to obtain the Galilean (super)gravity action in $D=3$ by using the Lie (super)algebras expansion method.
Although this method may give less physical insight than the procedures based on limits, it has the advantage of being
systematic and involving simpler calculations.

We have applied here our method to the $D=3$ case, but in principle it could be applied to higher dimensions, provided the
starting action is one that can be expressed as the integral over spacetime of a differential form constructed from the fields
and curvatures of a certain Lie (super)algebra. Since for $D>4$ the starting (super)gravity action will no longer be gauge invariant,
a question to be answered is to what extent the actions obtained by expansion are invariant under some symmetries of the expanded
algebra.

 The expansion method has recently been used in \cite{BIOR:19} to derive general actions for any $D$
 and $p$-brane\footnote{An example of string ($p=1$) action was found in \cite{GO:01}}, thus recovering
 some known examples of actions existing in the literature, and providing a way of reproducing others,
 such as Carrol gravity \cite{BGRRV:17,H:15}. This indicates that our method, at least in the bosonic case,
 can be applied when $D>3$. So it is natural to think that maybe this is also possible in the supersymmetric
 case. One potential problem is the fact that, in general, Poincar\'e supergravity
 with $N=2$ is required as the starting point. In $D=4$, for instance, the first order supergravity action
 contains not only the gauge fields of a centrally extended $N=2$ superPoincar\'e
 algebra, but also some auxiliary zero-forms that are needed to write in first order form the
 kinetic term of the gauge field associated with the central extension generator (see \cite{CDF:91}).
 So this problem may presumably be overcome by applying the expansion method to general free differential algebras,
 the gauge algebra being just an example.

Another difficulty is the local supersymmetry of the actions obtained by expansion. In \cite{BIOR:19},
it was shown that the actions do possess local symmetries corresponding to the generators of their algebras, but
the argument given there will not be applicable in the case of supersymmetry. Also, it is well known
that in the case of Poincar\'e supergravity the supersymmetry variations realize the supersymmetry algebra up
to field equations. It is not clear whether this is going to be the case when applying the expansion procedure.
One possible approach is to take as the starting point the action with auxiliary
fields that ensure the closure of the supersymmetry algebra off-shell but, also here,
the auxiliary fields do not correspond to gauge fields of a Lie algebra, so they
should be treated as zero forms of a free differential algebra.

\section{Acknowledgements}

This work has been partially supported by the grants
MTM2014-57129-C2-1-P from the MINECO (Spain), VA137G18 Spanish Junta de Castilla y Le\'on and
FEDER BU229P18.
Useful conversations with E. Bergshoeff, T. Ort\'{\i}n and  L. Romano are also appreciated.

\end{document}